\documentclass[preprint,preprintnumbers,showpacs,showkeys,amsmath,amssymb,nofootinbib,superscriptaddress]{revtex4}

\usepackage{color}
\usepackage{latexsym}
\usepackage{amsmath}
\usepackage{amssymb}
\usepackage{euscript}
\usepackage{pstricks}
\usepackage{graphics}
\usepackage{graphicx}
\usepackage{picture}

\newcommand{\bee}{\begin{equation}}
\newcommand{\eee}{\end{equation}}
\newcommand{\baa}{\begin{eqnarray}}
\newcommand{\eaa}{\end{eqnarray}}

\def\ni{\noindent}

\begin{document}
\title{High-derivatives and massive electromagnetic models
	in \\ the Lema\^{\i}tre-Tolman-Bondi spacetime}

\author{Rafael L. Fernandes}\email{rafael.fernandes@ifba.edu.br}
\affiliation{Instituto Federal de Educa\c{c}\~ao, Ci\^encia e Tecnologia
             da Bahia - Campus Juazeiro, Rodovia BA 210, S/N, Bairro
             Nova Juazeiro, 48918-900, Juazeiro, BA, Brazil}
\author{Everton M. C. Abreu}\email{evertonabreu@ufrrj.br}
\affiliation{Departamento de F\'{i}sica, Universidade Federal Rural do
             Rio de Janeiro, 23890-971, Serop\'edica, RJ, Brazil}
\affiliation{Departamento de F\'{i}sica, Universidade Federal de Juiz de
             Fora, 36036-330, Juiz de Fora, MG, Brazil}
\affiliation{Programa de P\'os-Gradua\c{c}\~ao Interdisciplinar em
             F\'isica Aplicada, Instituto de F\'{i}sica, Universidade
             Federal do Rio de Janeiro, 21941-972, Rio de Janeiro, RJ,
             Brazil}
\author{Marcelo B. Ribeiro}\email{mbr@if.ufrj.br}
\affiliation{Programa de P\'os-Gradua\c{c}\~ao Interdisciplinar em
             F\'isica Aplicada, Instituto de F\'{i}sica, Universidade
             Federal do Rio de Janeiro, 21941-972, Rio de Janeiro, RJ,
             Brazil}
\affiliation{Instituto de F\' isica, Universidade Federal do Rio de
            Janeiro, 21941-972, Rio de Janeiro, RJ, Brazil}
\date{\today}

\begin{abstract}
\ni The Maxwell electromagnetic theory embedded in an inhomogeneous
Lema\^{\i}tre-Tolman-Bondi (LTB) spacetime background was described
a few years back in the literature. However, terms concerning the mass
or high-derivatives were no explored. In this work we studied the
inhomogeneous spacetime effects on high-derivatives and massive
electromagnetic models. We used the LTB metric and calculated the
physical quantities of interest, namely the scale factor, density of
the eletromagnetic field and Hubble constant, for the Proca and
higher-derivative Podolsky models. We found a new singularity in both
models, and that the magnetic field must be zero in the Proca model.
\end{abstract}

\maketitle

\section{Introduction}

Homogeneity and isotropy, together with matter being treated as a specific
gas, are the basic ingredients of the Friedmann-Lema\^{\i}tre-Robertson-Walker
(FLRW) model. Since the pioneering work of E.\ Hubble, who suggested a
correlation between the observed redshifts and distances of 24 galaxies,
this is a very successfull model, being currently considered a very good
measure to describe our cosmos.

In the last decades a great effort was dedicated to understanding the local
inhomogeneities that occur in the Universe. Such efforts suggest that these
inhomogeneities may be related to the expansion of the Universe.
Inhomogeneities, as an alternative to dark energy, were first discussed in
Ref.\ \cite{Pascual} as a means of explaning the observational results of
the expansion of the Universe \cite{Rasanen,Chuang,Paranjape,Kai,Rasanen2,
Enqvist,Cosmai} without the need for postulating dark energy.
Inhomogeneities inside astronomical objects can define different
instability ranges, an effect that can describe distinct features of their
evolution and structures formation.

The current standard model of cosmology, the $\Lambda$-Cold Dark Matter
($\Lambda$CDM) model, is a homogeneous solutions of the FLRW Einstein's
field equations with only six free parameters that successfully accounts
for most cosmological data, especially the characteristics of the Cosmic
Microwave Background (CMB) and the structure formation on large scales
considered through the theory of cosmological perturbations in homogeneous
and isotropic background. However, in the last fifteen years ``standard''
inhomogeneous cosmological models that are generalizations of FLRW
cosmologies have been the subject of growing interest in astrophysics
community in order to investigate cosmological phenomena. Some authors
have demonstrated that inhomogeneous models with spherical symmetry and
dust source can be fitted to supernovae Ia (SNIa) data, as well as the
position of the first peak of the CMB. These models show that the
apparent accelerated expansion of the universe may not be a consequence
of the repulsive gravity due to dark energy, but rather the result of
inhomogeneities in the distribution of matter. In this context one needs
to mention that an important cosmological model to describe the
inhomogeneity universe is the Lema\^{\i}tre-Tolman-Bondi (LTB) spacetime
\cite{Bondi,Krasinski,Lemaitre,Ribeiro,Ribeiro1,Ribeiro2,Ribeiro3,Ribeiro4,
Nogueira}, which is a spatially inhomogeneous description of a spherically
symmetric distribution of dust matter in the Universe.

In the last few years several studies analyzing the scenario of
electrodynamics embedded in the isotropic and homogeneous FLRW gravitational
background models \cite{ciarcelluti} have been produced. Astrophysical
effects were explored together with an anisotropic expansion of the
Universe. The interesting result of polarized electromagnetic radiation
occurs when it travels through local anisotropic regions. Concerning the
inhomogeneities, in Ref.\ \cite{Fanizza} the authors investigated that the
inhomogeneity with electromagnetic field caused a new scale factor. The
propagation of photons was also affected, which is important phenomenum 
since most information obtained about the Universe is by means of photons.
Ref.\ \cite{ybr} studied the effects of Palatini $f(R)$ gravity together
with the so-called tilted observer on the dynamics of LTB spacetime
embedded in an electromagnetic field.

In this work we studied the effect of a high-derivative electromagnetic
field embedded in the inhomogeneous LTB geometry. We analyzed the Proca
and high-derivatives Podolsky electromagnetic models. The electrodynamic
contribution was inserted separately through the energy-momentum
tensor (EMT) of the respective models. For the Proca model, besides the
matter density there also are the electromagnetic contributions obtained
from the Lagrangian of the Proca model in curved space-time. An analysis
of the Proca model in curved space-time was carried out for a particular
case of interest by Bekenstein \cite{Bekenstein}. Here, we have analyzed
the electrodynamics effects in LTB cosmological model and calculate the
scale factor in LTB universe. We also computed the luminosity distance in
the presence of electromagnetic field.

Concerning the Podolsky electrodynamics effects in LTB cosmological model,
the study of Podolsky model in curved space-time was made in Ref.\
\cite{Cuzinatto} together with the analysis of the Bopp-Podolsky black
holes. Therefore, we have obtained the line element and the equations
that define the LTB model with Podolsky contributions.

This paper is structured as follows. In Section 2 we review the main
aspects of the LTB metric, where the Einstein tensor and the matter
contribution for the EMT are obtained. In Section 3 we analyze the
Proca model of electrodynamics in curved space-time, where the EMT and
the Maxwell-Proca equations in curved space-time are derived. In
Section 4 we solve the Einstein equations including the Proca
contribution. We obtain the scale factor for this model and examine
the inhomogeneities and luminosity distance. In section 5 we analyze
the Podolsky electrodynamics contributions for LTB model. Section 6
discusses the results and presents some final considerations.

\section{The LTB cosmological model: a brief review}

The LTB model depicts a self-gravitating spherically symmetric distribution
of inhomogeneous nondissipative dust cloud where the EMT can be written as
$T_{\mu\nu} = \epsilon (\tau,\rho)u_\mu u_\nu$ and $u_\mu = u_\mu (\tau,\rho)$
is the dust particle's four-velocity vector. The proper time is represented
by $\tau$ and the several shells are labeled by $\rho$, which helps in the
formation of the dust cloud.  As mentioned above, the LTB metric is the most
popular inhomogeneous cosmological model, since it is appropriate for both
small and large scale inhomogeneities, and also is the simplest inhomogeneous
solution of the Einstein equations.

In this section we will analyze the LTB model through the point of view of
the field equations of general relativity, as well as the matter contribution
concerning the EMT. In LTB cosmology, the line element is written as 
\begin{equation}
dS^2=dt^2-e^Adr^2-R^2\Big(d\theta^2+sin^2{\theta}d\phi^2\Big),
\label{1}
\end{equation}
where $A$ and $R$, unlike FLRW model, depend also on the $r$ coordinate,
namely, $A{\equiv}A(r,t)$ and $R{\equiv}R(r,t)$. Hence, the metric element
is given by
\begin{equation}
g_{\mu\nu}=\Big(1,-e^A,-R^2,-R^2sin^2\theta\Big),
\label{2}
\end{equation}
where
\begin{equation}
\sqrt{-g}=e^{A/2}R^2sin\theta.
\label{3}
\end{equation}
The spatially homogeneous FLRW metric is a special case of the metric in
Eq.\ \eqref{1}, where 
\bee
\label{2.1}
e^{A(r,t)} \,=\,\frac{a(t)}{\sqrt{1\,-\,kr^2}}
\eee
and $R(r,t) = a(t) r$, where $a(t)$ is the scale factor.

The Einstein equations, where $c=1$, are given by
\begin{equation}
G_{\mu\nu}=8{\pi}GT_{\mu\nu},
\label{4}
\end{equation}
where $G_{\mu\nu}$ is the Einstein tensor
\begin{equation}
G_{\mu\nu}=R_{\mu\nu}-\frac{1}{2}g_{\mu\nu}R,
\label{4.1}
\end{equation}
$G$ is the gravitational constant and $T_{\mu\nu}$ is the EMT. So, we
can make use of the line element in Eq. $(\ref{1})$ to obtain the
non-zero components of the Einstein tensor, 
\begin{align}
G^0_{\;0}&=\Big(-2\frac{R''}{R}+\frac{R'A'}{R}-\frac{R'^2}{R}\Big)e^{-A}+
	\Big(\frac{\dot{R}\dot{A}}{R}+\frac{\dot{R}^2}{R^2}+\frac{1}{R^2}
	\Big)\,\,,\nonumber\\
G^1_{\;1}&=\frac{\dot{R}^2}{R^2}+2\frac{\ddot{R}}{R}-e^{-A}\frac{R'^2}
	{R^2}+\frac{1}{R^2}\,\,,\nonumber\\
G^2_{\;2}&=G^3_{\;3}=\Big(-\frac{R''}{R}+\frac{A'R'}{2R}\Big)e^{-A}+
	\frac{\ddot{R}}{R}+\frac{\ddot{A}}{2}+\frac{\dot{A}^2}{4}-
	\frac{\dot{R}\dot{A}R}{2}\,\,,\nonumber\\
G^1_{\;0}&=e^{-A}\Big(2\frac{\dot{R}'}{R}-\frac{\dot{A}R'}{R}\Big)\,\,.
\label{5}
\end{align}

In LTB cosmology, the EMT is the zero pressure diagonal perfect fluid,
$(\rho,0,0,0)$, where $\rho$ is the mass density. So, the matter
contribution for the EMT $T_{00}^M$ is
\begin{equation}
T_{00}^M=\rho.
\label{5.1}
\end{equation}

The electromagnetic contribution for the LTB cosmology will be obtained
by adding on the right side of the Einstein equations. So, we have for
the EMT, for both the Proca model and for the Podolsky model, that
\begin{equation}
T_{\mu\nu}=T_{\mu\nu}^M+T_{\mu\nu}^{EM},
\label{5.2}
\end{equation}
where $T_{\mu\nu}^{EM}$ is the electromagnetic contribution for the EMT.
Therefore, determining  $T_{\mu\nu}^{EM}$, then we have the Einstein equation.

\section{The curved spacetime Proca model}

The Proca model of electrodynamics is defined by the following Lagrangian 
\begin{equation}
\mathcal{L}_{Proca}=-\frac{1}{4}F_{\mu\nu}F^{\mu\nu}+\frac{1}{2}
m^2A_{\mu}A^{\mu},
\label{6}
\end{equation}
where $m$ is the mass of electromagnetic field, $A_\mu$ is the
four-potential and $F_{\mu\nu}$ is the electromagnetic tensor that, in
curved space-time, is defined by $F_{\mu\nu}=\nabla_{\mu}
A_\nu-\nabla_{\mu}A_\nu$, where $\nabla$ is the covariant derivative
given by
\begin{equation}
\nabla_{\gamma}F_{\alpha\beta}=\partial_{\gamma}F_{\alpha\beta}-
\Gamma^{\rho}_{\gamma\alpha}F_{\rho\beta}-\Gamma^{\rho}_{\gamma\beta}
F_{\alpha\rho}.
\label{6.1}
\end{equation}
Moreover, the electromagnetic field and its dual can be obtained from
the following expressions,
\begin{align}
F_{\mu\nu}&=u_{\mu}E_\nu-u_{\nu}E_\mu+\epsilon_{\mu\nu\alpha\beta}
	B^\alpha{u}^\beta\nonumber\\
^*F^{\mu\nu}&=\frac{1}{2}\epsilon^{\mu\nu\alpha\beta}F_{\alpha\beta}
\label{6.2}
\end{align}
and the Maxwell-Proca equations in curved spacetime are
\begin{align}
\nabla_{\nu}F^{\mu\nu}+m^2A^{\mu}&=4\,{\pi}J^\mu\nonumber\\
{\nabla_{\nu}}^*F^{\mu\nu}&=0,
\label{6.3}
\end{align} 
where $J^\mu=(\rho,\vec{J})$ is the electromagnetic four-current density
and $^*F^{\mu\nu}$ is the dual electromagnetic tensor. 

From Eq. $(\ref{6})$, the action for the Proca model is 
\begin{equation}
S=\frac{1}{16\pi}{\int}d^4x\sqrt{-g}\Big(-R-4\mathcal{L}\Big),
\label{7}
\end{equation}
namely,
\begin{equation}
S=\frac{1}{16\pi}{\int}d^4x\sqrt{-g}\Big(-R+F_{\mu\nu}F^{\mu\nu}-2m^2A_{\mu}A^{\mu}\Big).
\label{7.1}
\end{equation}
and, from the EMT definition
\begin{equation}
T_{\mu\nu}=\frac{2}{\sqrt{-g}}\frac{\delta{S}}{\delta{g^{\mu\nu}}},
\label{8}
\end{equation}
the Proca EMT in curved spacetime is 
\begin{equation}
T_{\mu\nu}=\frac{1}{4\pi}\Big[F_{\mu}^{\rho}F_{\nu\rho}+m^2A_{\mu}
A_{\nu}+g_{\mu\nu}\Big(\frac{1}{4}F_{\alpha\beta}F^{\alpha\beta}-
\frac{1}{2}m^2A_{\alpha}A^{\alpha}+J^\alpha\tilde{A}_\alpha\Big)\Big],
\label{9}
\end{equation}
In this work, we neglect the coupling term $J^\alpha\tilde{A}_\alpha$ 
because most of the matter is electrically neutral. Thus, any fluctuations that may appear can be ignored. The trace of the energy-momentum tensor is: 
\begin{equation}
T=-\frac{5}{8\pi}m^2A_{\mu}A^{\mu},
\label{10}
\end{equation}
i.e., unlike Maxwell's case, the tensor trace is not zero.

Using relations in Eq. $(\ref{9})$, we find that the non-zero components
of the EMT are
\begin{align}
T_{00}&=e^A(E^{(1)})^2+R^2(E^{(2)})^2+R^2sin^2(E^{(3)})^2+\frac{1}{4}F^{\alpha\beta}F_{\alpha\beta}+\frac{m^2}{4\pi}A_0^2-\frac{1}{8\pi}m^2A_{\alpha}A^\alpha,\nonumber\\
T_{11}&=-e^{2A}(E^{(1)})^2-e^AR^2sin^2(B^{(3)})^2+e^AR^2(B^{(2)})^2+\frac{e^A}{4}F^{\alpha\beta}F_{\alpha\beta}+\frac{m^2}{4\pi}A_1^2-\frac{1}{8\pi}e^Am^2A_{\alpha}A^\alpha,\nonumber\\
T_{22}&=-R^4E^{(2)}+R^4sin^2\theta(B^{(3)})^2+e^AR^2(B^{(1)})^2-\frac{R^2}{4}F^{\alpha\beta}F_{\alpha\beta}+\frac{m^2}{4\pi}A_2^2-\frac{1}{8\pi}g_{22}m^2A_{\alpha}A^\alpha\nonumber\\
T_{33}&=-R^4E^{(2)}+R^4sin^2\theta(B^{(3)})^2+e^AR^2(B^{(1)})^2-\frac{R^2}{4}F^{\alpha\beta}F_{\alpha\beta}+\frac{m^2}{4\pi}A_3^2-\frac{1}{8\pi}g_{33}m^2A_{\alpha}A^\alpha\nonumber\\
T_{10}&=e^{A/2}R^2sin\theta(E^{(3)}B^{2)}-E^{(2)}B^{(3)})+\frac{m^2}{4}A_1A_0,\nonumber\\
T_{20}&=e^{A/2}R^2sin\theta(E^{(1)}B^{(3)}-E^{(3)}B^{(1)})+\frac{m^2}{4}A_2A_0,\nonumber\\
T_{30}&=e^{A/2}R^2sin\theta(E^{(2)}B^{(1)}-E^{(1)}B^{(2)})+\frac{m^2}{4}A_3A_0,\nonumber\\
T_{12}&=-e^{A}R^2(E^{(1)}E^{(2)}+B^{(1)}B^{(2)})+\frac{m^2}{4}A_1A_2,\nonumber\\
T_{13}&=-e^{A}R^2sin^2\theta(E^{(1)}E^{(3)}+B^{(1)}B^{(3)})+\frac{m^2}{4}A_1A_3,\nonumber\\
T_{23}&=-R^4sin^2\theta(E^{(2)}E^{(3)}-B^{(2)}B^{(3)})+\frac{m^2}{4}A_2A_3.
\label{11}
\end{align}

The next step is introduce Eqs. $(\ref{11})$ into the Einstein equations in Eq. $(\ref{4})$. We will do this in the next section and we will analyze the effects of the Proca model on the LTB model.

\section{LTB scenario Proca electrodynamics}

Now that we have the Proca's contribution in hand, we must analyze the terms of the Einstein tensor to obtain Einstein's equations. As we can see from Eq. $(\ref{5})$, the only non-zero off-diagonal term in $G_{\mu\nu}$ is $G_{10}$.   The only way to satisfy this condition is to impose that
\begin{equation}
T_{02}=T_{03}=T_{12}=T_{13}=T_{23},
\label{12}
\end{equation}
and, consequently
\begin{equation}
\begin{cases}A_2=A_3=0\\
F_{02}=F_{03}=F_{12}=F_{13}=F_{23}=0.\end{cases}
\label{13}
\end{equation}

Taking into account the last conditions, the electromagnetic fields are
\begin{align}
E^\mu=&(0,E(t,r),0,0),\nonumber\\
B^\mu=&(0,0,0,0),
\label{13.1}
\end{align}
as we can see, as consequence of conditions in Eq. $(\ref{13})$, we have
that the $B_\mu$ field is zero. The only contribution of electromagnetic
field is by $E_\mu$ field. Now, imposing these same conditions in equations
$(\ref{11})$, the energy-momentum components are
\begin{align}
T_{00}&=\frac{1}{2}e^{A}E^2+\frac{m^2}{8\pi}A_0^2-\frac{m^2}{8\pi}g^{11}A_1^2,\nonumber\\
T_{11}&=\frac{1}{2}e^{2A}E^2-\frac{m^2e^A}{8\pi}\Big(A_0^2+A_1^2\Big),\nonumber\\
T_{22}&=T_{33}=-\frac{R^2e^A}{2}E^2-g_{33}\frac{1}{8\pi}m^2A_{0}^2,\nonumber\\
T_{01}&=\frac{1}{8\pi}\Big(m^2A_0A_1\Big).
\label{14}
\end{align}

In order to write $A$ in function of $R$, we impose that $A_1=0$, that is $A_\mu=(A_0,0,0,0)$. So, the equations $(\ref{11})$ turns
\begin{align}
T_{00}&=\frac{1}{2}e^{A}E^2+\frac{m^2}{8\pi}A_0^2,\nonumber\\
T_{11}&=\frac{1}{2}e^{2A}E^2-\frac{m^2e^A}{8\pi}\Big(A_0^2\Big),\nonumber\\
T_{22}&=T_{33}=-\frac{R^2e^AE^2}{2}-g_{33}\frac{1}{8\pi}m^2A_{0}^2,\nonumber\\
T_{01}&=0.
\label{15}
\end{align}
and the Einstein equations stay with matter e electromagnetic contribution
in EMT are 
\begin{align}
G^0_0&=\Big(-2\frac{R''}{R}+\frac{R'A'}{R}-\frac{R'^2}{R}\Big)e^{-A}+\Big(\frac{\dot{R}\dot{A}}{R}+\frac{\dot{R}^2}{R^2}+\frac{1}{R^2}\Big)=8{\pi}G\Big[\rho_m+\frac{1}{2}e^{A}E^2+\frac{m^2}{8\pi}A_0^2\Big],\nonumber\\
G^1_1&=-\frac{\dot{R}^2}{R^2}+2\frac{\ddot{R}}{R}-e^{-A}\frac{R'^2}{R^2}+\frac{1}{R^2}=-8{\pi}G\Big[\frac{1}{2}e^{A}E^2-\frac{m^2}{8\pi}\Big(A_0^2\Big)\Big],\nonumber\\
G^2_2&=G^3_3=\Big(-\frac{R''}{R}+\frac{A'R'}{2R}\Big)e^{-A}+\frac{\ddot{R}}{R}+\frac{\ddot{A}}{2}+\frac{\dot{A}^2}{4}-\frac{\dot{R}\dot{A}R}{2}=-g^{33}\frac{R^2e^AE^2}{2}-\frac{1}{8\pi}m^2A_{0}^2,\nonumber\\
G^1_0&=e^{-A}\Big(2\frac{\dot{R}'}{R}-\frac{\dot{A}R'}{R}\Big)=0.
\label{16}
\end{align}

The last equation of Eqs.\ $(\ref{16})$ allows us as to write $A$ in terms
of $R'$ such as
\begin{equation}
e^A=\frac{R'^2}{1-k(r)}
\label{17}
\end{equation} 
where $k(r)$ is an arbitrary function. So, the line element is
\begin{equation}
dS^2=dt^2-\frac{R'^2}{1-k(r)}dr^2-R^2\Big(d\theta^2+sin^2{\theta}d\phi^2\Big)\,\,.
\label{17.1}
\end{equation}
Moreover, we can rewrite the Einstein equations for the $(00)$ and $(11)$ components like
\begin{equation}
\frac{\dot{R}^2+k}{R^2}+\frac{2\dot{R}\dot{R}'+k'}{RR'}=8{\pi}G\,\bigglb[\rho_m+\frac{R'^2E^2}{(1-k)}+m^2A_0^2-J^{\alpha}A_\alpha\biggrb],
\label{18.1}
\end{equation}
\begin{equation}
\frac{\dot{R}^2+2R\ddot{R}+k}{R^2}=8{\pi}G\bigglb[\frac{R'^2E^2}{(1-k)}+m^2A_0^2-J^{\alpha}A_\alpha\biggrb].
\label{18.2}
\end{equation}

On the other hand, the non-zero components of the Maxwell-Proca equations in Eqs. $(\ref{6.2})$ are
\begin{equation}
\partial_1\Big(e^{A/2}R^2F^{01}+m^2e^{A/2}R^2A^0\Big)=4{\pi}e^{A/2}R^2J^0,
\label{20.1}
\end{equation}
\begin{equation}
\partial_0\Big(e^{A/2}R^2F^{10}\Big)=4{\pi}e^{A/2}R^2J^1,
\label{20.2}
\end{equation}
\begin{equation}
\partial_2E^{(1)}=0,
\label{20.3}
\end{equation}
\begin{equation}
\partial_3E^{(1)}=0.
\label{20.4}
\end{equation}

So, we can write the electric field $E(t,r)$ using the integration of Eq. $(\ref{20.1})$, as
\begin{equation}
E(t,r)=\frac{\epsilon(t)+\xi(t,r)}{e^{A/2}R^2},
\label{21}
\end{equation}
where 
\begin{equation}
\xi(t,r){\equiv}4\pi\int_0^re^{A/2}R^2J^0d\bar{r}-m^2e^{A/2}R^2A^0,
\label{22}
\end{equation}
and $\epsilon(t)$ is the constant of integration.   Now, from Eq.\
$(\ref{20.2})$, using the fact that the charged matter belongs to the
comoving matter, namely, the four-current appears as $J^\alpha=(J^0,\vec{0})$
\begin{equation}
\dot{\epsilon}+\dot{\xi}=0,
\label{23}
\end{equation}
which leads us to
\begin{equation}
E(t,r)=\frac{\epsilon_0+\xi_0(r)}{e^{A}R^4}.
\label{24}
\end{equation}

Now, defining 
\begin{equation}
\sigma(r)\,{\equiv}\,4{\pi}Ge^AR^4(E^2+2e^{-A}m^2A^2_0),
\label{25}
\end{equation}
multiplying the equation $(\ref{18.2})$ by $\dot{R}$, we obtain that
\begin{equation}
\dot{R}^3+2A\dot{R}\ddot{R}=-k\dot{R}+\sigma\frac{\dot{R}}{R^2},
\label{26}
\end{equation}
or yet, using the fact that $\partial_0(R\dot{R})=2A\dot{R}\ddot{R}$, we have
\begin{align}
R\dot{R}&=-k(r)R+\alpha(r)+\int\frac{\sigma(r)}{R^2}\dot{R}dt,\nonumber\\
\dot{R}^2&={-k(r)}+\frac{\alpha(r)}{R}-\frac{\sigma(r)}{R^2},
\label{27}
\end{align}
and using Eq. $(\ref{18.1})$ we can write
\begin{align}
&8\pi{G}\Big(\rho+\frac{R'E^2}{(1-k)}+m^2A_0^2\Big)\nonumber\\
&=\frac{k'}{RR'}+\frac{\alpha(r)}{R^3}-\frac{\sigma(r)}{R^4}+\frac{2}{RR'}\Big[\Big(-k(r)-\frac{\alpha(r)}{R}-\frac{\sigma(r)}{R^2}\Big)\Big(\frac{\alpha{R'}}{R^2}+2\frac{\sigma(r)R'}{R^3}\Big)\Big]\nonumber\\
\label{27.1}
\end{align}
Eqs. $(\ref{17.1})$, $(\ref{27})$ and $(\ref{27.1})$ define the LTB model with the Proca electrodynamics contributions. The singularities arising from $R=0$, $R'=0$ and $k=1$. The $R=0$ singularity we interpreted as the Big Bang singularity, and the $R'=0$ singularity as the shell cross singularity. The last singularity, $k=1$, that occurs if $R'\neq{0}$, come from the Proca contribution.

Now, defining
\begin{align}
-k(r)\,{\equiv}\,H_0^2(r)\Omega_k(r)R_0^2(r),\nonumber\\
\alpha(r)\,{\equiv}\,H_0^2(r)\Omega_m(r)R_0^3(r),\nonumber\\
-\sigma(r)\,{\equiv}\,H_0^2(r)\Omega_\sigma(r)R_0^4(r),
\label{28}
\end{align}
where the Hubble constant is defined as
\begin{equation} 
H(t,r)=\frac{\dot{R}}{R}\,\,.
\label{28.1}
\end{equation}
We have imposed the boundary values at $t_0$ through $A_0(r){\equiv}A(t_0,r)$, $H_0(r){\equiv}H(t_0,r)$. Moreover, $\Omega_k$, $\Omega_m$ and $\Omega_\sigma$ are subjected to the constraint
\begin{equation}
\Omega_k(r)+\Omega_\sigma(r)+\Omega_m(r)=1\,\,.
\label{29}
\end{equation}

Therefore, we can write Eq. $(\ref{27})$ as
\begin{equation}
\frac{\dot{R}}{R}=H_0(r)\Big[\Omega_k(r)\Big(\frac{R_0}{R}\Big)^2+\Omega_m(r)\Big(\frac{R_0}{R}\Big)^3+\Omega_\sigma(r)\Big(\frac{R_0}{R}\Big)^4\Big]^{1/2}
\label{30}
\end{equation}
and the Hubble constant is computed by
\begin{equation}
-H_0(r)t=-\int_{\frac{R}{R_0}}^1\frac{dx}{\sqrt{\Omega_k(r)+\Omega_\sigma(r)x^{-1}+\Omega_m(r)x^{-2}}}
\label{31}
\end{equation}
and, considering the solution using that $\dot{R}>0$, we have
\begin{equation}
R(t,r)=R_0(r)\bigglb[\frac{\Omega_\sigma(r)}{\Omega_m(r)}+\frac{\Omega_m(r)\Omega_\sigma^2(r)}{M(t,r)}+\frac{M(t,r)}{\Omega_m^3(r)}\biggrb]\,\,,
\label{32}
\end{equation}
where 
\begin{align}
&M(t,r)=\Omega_m^2(r)\bigglb(\frac{N+2\Omega_\sigma^3(r)+\sqrt{N^2+4N\Omega_\sigma^6(r)}}{2}\biggrb)^{1/3}\,\,,\\
&N(t,r)=\bigglb[\frac{3\Omega_m(r)H_0(r)t}{2}+\Omega_m(r)-2\Omega_\sigma(r)\biggrb]^2-4\Omega_\sigma^3(r)\,\,,
\label{33}
\end{align}
where $R_0(r)$ corresponds to the current shape of the scale factor, $H_0(r)$ is the current value of the Hubble constant in each point and $\Omega_\sigma(r)$ is the density of the electromagnetic field, that, because of Proca's contributions, is purely electric.

Remember that the density of the electromagnetic field is given by
\begin{equation}
\Omega_\sigma=-\frac{\sigma(r)}{H_0^2R_0^4}
\label{33.1}
\end{equation}
where
\begin{equation}
\sigma(r)\,{\equiv}\,4{\pi}Ge^AR^4\Big(E^2+2e^{-A}m^2A^2_0\Big)\,\,,
\end{equation}
and the Proca contribution for the density of the electromagnetic field is the mass term $-2e^{-A}m^2A^2_0$.

Now, let us examine the inhomogeneities and luminosity distance for the Proca electrodynamics perspective. 
Therefore, the geodesic requires that $d\theta=d\phi=0$.  Moreover, since light always travels along null geodesics, we have $dS^2=0$. So, using Eq. $\eqref{17.1}$, we have
\begin{equation}
dt^2=\frac{R'^2}{1+2k(r)}dr^2
\label{34}
\end{equation}
so that
\begin{equation}
\frac{dt}{du}=\frac{R'}{\sqrt{1+2k(r)}}\frac{dr}{du}\,\,.
\label{35}
\end{equation}

Consider two light rays with solutions of Eq. $(\ref{35})$ given by $t_1 = t(u)$ and $t_2 = t(u) + \lambda(u)$. 
Substituting these two into Eq. $(\ref{35})$ we obtain
\begin{align}
\frac{dt_1}{du}&=-\frac{dr}{du}\frac{R'}{\sqrt{1+2k(r)}},\nonumber\\
\frac{dt_2}{du}&=-\frac{dr}{du}\frac{R'+\dot{R}\lambda}{\sqrt{1+2k(r)}},\nonumber\\
\frac{d\lambda}{du}&=-\frac{dr}{du}\frac{\dot{R}'\lambda}{\sqrt{1+2k(r)}}.
\label{36}
\end{align}

Differentiating the definition of the redshift, $z\,{\equiv}\,\Big[\lambda(0)-\lambda(u)\Big]/\lambda(u)$, we have,
\begin{equation}
\frac{dz}{du}=\frac{dr}{du}\frac{(1+z)\dot{R}'}{\sqrt{1+2k(r)}}
\label{37}
\end{equation}
and, consequently,
\begin{align}
\frac{dt}{dz}&=-\frac{R'}{(1+z)\dot{R}'}\,\,,\nonumber\\
\frac{dr}{dz}&=\sqrt{\frac{1-k(r)}{(1+z)\dot{R}'}}\,\,,
\label{38}
\end{align}
which determine the relation between the coordinates and the observable redshift.

For the last equation, using the expression for $k(r)$
\begin{equation}
k(r){\equiv}H_0^2\Omega_kA_0^2
\label{39}
\end{equation}
we find
\begin{equation}
\frac{dr}{dz}=\sqrt{\frac{1+H_0^2(r)(1-\Omega_m(r)+\Omega\sigma(r))R_0^2(r)}{(1+z)\dot{R}'(t,r)}}\,\,.
\label{40}
\end{equation}
Moreover, the relation between the redshift and the energy flux F, defined as $d_L{\equiv}\sqrt{L/(a\pi{F})}$, where $L$ is the total power distance radiated by the source, is
\begin{equation}
d_L(z)=(1-z)^2R(r(z),t(z))
\label{41}
\end{equation}
and the angular distance diameter is given by
\begin{equation}
d_A(z)=R(r(z),t(z))
\label{42}
\end{equation}
which is a direct relation to the scale factor as a function of the redshift.

\section{LTB geometry Podolsky electrodynamics}
The Podolsky electrodynamics in curved space-time is given by
\begin{equation}
\mathcal{L}_{Pod}=-\frac{1}{4}F^{\alpha\beta}F_{\alpha\beta}+\frac{a^2}{2}\nabla_{\beta}F^{\alpha\beta}\nabla_{\gamma}F_\alpha^{\;\gamma}+\frac{b^2}{2}\nabla_{\beta}F^{\alpha\gamma}\nabla^{\beta}F_{\alpha\gamma}
\label{43}
\end{equation}
or,
\begin{align}
\mathcal{L}_{Pod}=&-\frac{1}{4}F^{\alpha\beta}F_{\alpha\beta}+\frac{a^2+2b^2}{2}\nabla_{\beta}F^{\alpha\beta}\nabla_{\gamma}F_\alpha^{\;\gamma}\nonumber\\
&+\frac{b^2}{2}\Big(R_{\sigma\beta}F^{\sigma\alpha}F_\alpha^{\;\beta}+R_{\alpha\sigma\beta\gamma}F^{\sigma\gamma}F^{\alpha\beta}\Big)
\label{44}
\end{align}
where $R_{\sigma\beta}$ is the Ricci tensor.

The Einstein-Podolsky action is
\begin{equation}
S=\frac{1}{16\pi}{\int}d^4x\sqrt{-g}\Big[-R+4\mathcal{L}_{Pod}\Big]
\label{45}
\end{equation}
which leads us, from the variation with respect to $A_\mu$, to the Einstein-Podolsky equation
\begin{equation}
\nabla_\nu\Big[F^{\mu\nu}-(a^2+2b^2)H^{\mu\nu}+2b^2S^{\mu\nu}\Big]=0
\label{46}
\end{equation}
where
\begin{align}
H^{\mu\nu}&\,{\equiv}\,\nabla^{\mu}K^\nu-\nabla^{\nu}K^\mu\nonumber\\
S^{\mu\nu}&\,{\equiv}\,F^{\mu\sigma}R_\sigma^\nu-F^{\nu\sigma}R_\sigma^\mu+2R^{\mu\;\nu}_{\;\sigma\;\beta}{F^{\beta\sigma}}
\label{47}
\end{align}
and
\begin{equation}
K^\mu\,{\equiv}\,\nabla_{\nu}F^{\mu\nu}.
\label{48}
\end{equation}

The EMT for the Podolsky electrodynamics in curved space-time is given by
\begin{align}
T_{\mu\nu}&=\frac{1}{4\pi}\Big[F_{\mu\sigma}F^\sigma_{\;\nu}+\frac{1}{4}g_{\mu\nu}F^{\alpha\beta}F_{\alpha\beta}\Big]\nonumber\\
&+\frac{a^2}{4\pi}\Big[g_{\mu\nu}F_\beta^{\;\gamma}\nabla_{\gamma}K^\beta+\frac{1}{2}g_{\mu\nu}K^{\beta}K_{\beta}+2F_{(\mu}^\alpha\nabla_{\nu)}K_\alpha-2F_{(\mu}^\alpha\nabla_{\alpha}K_{\nu)}-K_\mu{K_\nu}\Big]\nonumber\\
&+\frac{b^2}{2\pi}\Big[\frac{1}{4}g_{\mu\nu}\nabla^{\beta}F^{\alpha\gamma}\nabla_{\beta}F_{\alpha\gamma}+F^\gamma_{\;(\mu}\nabla^\beta\nabla_{\beta}F_{\nu)\gamma}
+F^{\beta}_{(\mu}\nabla_\beta\nabla_{\nu)}-\nabla_{\beta}\Big(F_\gamma^{\;\beta}\nabla_{(\mu}F_{\nu)}^{\;\gamma}\Big)\Big]
\label{49}
\end{align}
and the trace of the EMT is
\begin{equation}
T=\frac{a^2+2b^2}{4\pi}K^{\mu}K_\mu+\frac{b^2}{4\pi}\Big[2\nabla_\beta\Big(F^{\gamma\alpha}\nabla^{\beta}F_{\alpha\gamma}\Big)+F^{\nu\mu}S_{\mu\nu}\Big]
\label{50}
\end{equation}
Now, using the metric element Eq. $(\ref{1})$, the non-null elements for the energy-momentum tensor are
\begin{align}
T^0_{\;0}&=e^A(E^{(1)})^2+R^2(E^{(2)})^2+R^2sin^2(E^{(3)})^2+\frac{1}{4}F^{\alpha\beta}F_{\alpha\beta}\nonumber\\
&+\frac{a^2}{4\pi}\Big[F_\beta^{\;\gamma}\nabla_{\gamma}K^\beta-\frac{K_0K^0}{2}+\frac{1}{2}K^iK_i+2F^{0i}\nabla_0K_i-2F^{0i}\nabla_iK_0\Big]\nonumber\\
&+\frac{b^2}{2\pi}\Big[-\frac{1}{4}\nabla^{\beta}F^{\alpha\gamma}\nabla_{\beta}F_{\alpha\gamma}+F^{i0}\nabla^\beta\nabla_{\beta}F_{0i}+F_{0i}\nabla_\beta\nabla^0{F}^{\beta{i}}-\nabla_\beta\Big(F_i^\beta\nabla^0{F}_0^i\Big)\Big]
\label{51}
\end{align}
\begin{align}
T^1_{\;1}&=e^A(E^{(1)})^2-R^2sin^2(B^{(3)})^2-R^2(B^{(2)})^2+\frac{1}{4}F^{\alpha\beta}F_{\alpha\beta}\nonumber\\
&+\frac{a^2}{4\pi}\Big[F_\beta^{\;\gamma}\nabla_{\gamma}K^\beta-\frac{1}{2}K^{\beta}K_{\beta}-2e^{-A}F_1^\alpha\nabla_{1}K_{\alpha}+2e^{-A}F_1^\alpha\nabla_{\alpha}K_1+K^1K_1\Big]\nonumber\\
&+\frac{b^2}{2\pi}\Big[-\frac{1}{4}\nabla^{\beta}F^{\alpha\gamma}\nabla_{\beta}F_{\alpha\gamma}+F^{\gamma{1}}\nabla^\beta\nabla_{\beta}F_{1\gamma}+F_\gamma^{1}\nabla_\beta\nabla_1{F}^{\beta\gamma}-\nabla_\beta\Big(F_{\gamma}^\beta\nabla_1{F}_1^\gamma\Big)\Big]
\label{52}
\end{align}
\begin{align}
T^2_{\;2}&= R^2E^{(2)}-R^2sin^2\theta(B^{(3)})^2-e^A(B^{(1)})^2+\frac{1}{4}F_{\alpha\beta}F^{\alpha\beta}\nonumber\\
&+\frac{a^2}{4\pi}\Big[F_\beta^{\;\gamma}\nabla_{\gamma}K^\beta+\frac{1}{2}K^\beta_\beta-F^{2\alpha}\nabla_2K_\alpha-2F^{2\gamma}\nabla_{\gamma}K_2-K^2K_2\Big]\nonumber\\
&+\frac{b^2}{2\pi}\Big[-\frac{1}{4}\nabla^{\beta}F^{\alpha\gamma}\nabla_{\beta}F_{\alpha\gamma}+F^\gamma_2\nabla^\beta\nabla_\beta{F}^2_\gamma+F_{\gamma{2}}\nabla_\beta\nabla^2F^{\beta\gamma}-\nabla_\beta\Big(F_\gamma^\beta\nabla_{2}F_{2}^\gamma\Big)\Big]
\label{53}
\end{align}
\begin{align}
T^3_{\;3}&=\frac{1}{4}g^{33}\Big[-R^4E^{(2)}+R^4sin^2\theta(B^{(3)})^2+e^AR^2(B^{(1)})^2-\frac{R^2}{4}F^{\alpha\beta}F_{\alpha\beta}\Big]\nonumber\\
&+\frac{a^2}{4\pi}\Big[g_{33}F_\beta^{\;\gamma}\nabla_{\gamma}K^\beta+\frac{1}{2}g_{33}K^\beta_\beta-F^{3\alpha}\nabla_3K_\alpha-2F^{3\gamma}\nabla_{\gamma}K_3-K^3K_3\Big]\nonumber\\
&+\frac{b^2}{2\pi}\Big[-\frac{1}{4}\nabla^{\beta}F^{\alpha\gamma}\nabla_{\beta}F_{\alpha\gamma}+F^\gamma_3\nabla^\beta\nabla_\beta{F}^3_\gamma+F_{\gamma{3}}\nabla_\beta\nabla^3F^{\beta\gamma}-\nabla_\beta\Big(F_\gamma^\beta\nabla_{3}F_{3}^\gamma\Big)\Big]
\label{54}
\end{align}
\begin{align}
T^1_{\;0}&=e^{A/2}R^2sin\theta\Big(E^{(3)}B^{2)}-E^{(2)}B^{(3)}\Big)+\frac{a^2}{4\pi}\Big[2F_{(1}^\alpha\nabla_{0)}K_\alpha-2F_{(1}^\alpha\nabla_{\alpha}K_{0)}\Big]\nonumber\\
&+\frac{b^2}{2\pi}\big[F^\gamma_{(1}\nabla_ \alpha\nabla_ {\beta}F_{0)\gamma}F^{\beta\alpha}-\nabla_{\beta}F_\gamma^\beta\nabla_{(1}F_{0)}^\gamma+\nabla_\beta{F}_\gamma^\beta\nabla_{(1}F_{0)}^\gamma-F_\gamma^\beta\nabla_\beta\nabla_{(1}F_{0)}^\gamma\Big]
\label{55}
\end{align}
\begin{align}
T^2_{\;0}&=g^{22}\Big\{e^{A}R^2sin\theta\Big(E^{(3)}B^{(2)}-E^{(2)}B^{(3)}\Big)+\frac{a^2}{4\pi}\Big[2F_{(2}^\alpha\nabla_{0)}K_\alpha-2F_{(2}^\alpha\nabla_{\alpha}K_{0)}\Big]\nonumber\\
&+\frac{b^2}{2\pi}\big[F^\gamma_{(2}\nabla_\alpha\nabla_{\beta}F_{0)\gamma}F^{\beta\alpha}-\nabla_{\beta}F_\gamma^\beta\nabla_{(2}F_{0)}^\gamma+\nabla_\beta{F}_\gamma^\beta\nabla_{(2}F_{0)}^\gamma-F_\gamma^\beta\nabla_\beta\nabla_{(2}F_{0)}^\gamma\Big]\Big\}
\label{56}
\end{align}
\begin{align}
T^3_{\;0}&=g^{33}\Big\{e^{A}R^2sin\theta\Big(E^{(3)}B^{(2)}-E^{(2)}B^{(3)}\Big)+\frac{a^2}{4\pi}\Big[2F_{(3}^\alpha\nabla_{0)}K_\alpha-2F_{(3}^\alpha\nabla_{\alpha}K_{0)}\Big]\nonumber\\
&+\frac{b^2}{2\pi}\big[F^\gamma_{(3}\nabla_\alpha\nabla_{\beta}F_{0)\gamma}F^{\beta\alpha}-\nabla_{\beta}F_\gamma^\beta\nabla_{(3}F_{0)}^\gamma+\nabla_\beta{F}_\gamma^\beta\nabla_{(3}F_{0)}^\gamma-F_\gamma^\beta\nabla_\beta\nabla_{(3}F_{0)}^\gamma\Big]\Big\}
\label{57}
\end{align}
\begin{align}
T^1_{\;2}&=g^{11}\Big\{e^{A}R^2sin\theta\Big(E^{(3)}B^{(2)}-E^{(2)}B^{(3)}\Big)+\frac{a^2}{4\pi}\Big[2F_{(1}^\alpha\nabla_{2)}K_\alpha-2F_{(1}^\alpha\nabla_{\alpha}K_{2)}\Big]\nonumber\\
&+\frac{b^2}{2\pi}\big[F^\gamma_{(1}\nabla_\alpha\nabla_{\beta}F_{2)\gamma}F^{\beta\alpha}-\nabla_{\beta}F_\gamma^\beta\nabla_{(1}F_{2)}^\gamma+\nabla_\beta{F}_\gamma^\beta\nabla_{(1}F_{2)}^\gamma-F_\gamma^\beta\nabla_\beta\nabla_{(1}F_{2)}^\gamma\Big]\Big\}
\label{58}
\end{align}
\begin{align}
T^1_{\;3}&=g^{11}\Big\{e^{A}R^2sin\theta\Big(E^{(3)}B^{(2)}-E^{(2)}B^{(3)}\Big)+\frac{a^2}{4\pi}\Big[2F_{(1}^\alpha\nabla_{3)}K_\alpha-2F_{(1}^\alpha\nabla_{\alpha}K_{3)}\Big]\nonumber\\
&+\frac{b^2}{2\pi}\big[F^\gamma_{(1}\nabla_\alpha\nabla_{\beta}F_{3)\gamma}F^{\beta\alpha}-\nabla_{\beta}F_\gamma^\beta\nabla_{(1}F_{3)}^\gamma+\nabla_\beta{F}_\gamma^\beta\nabla_{(1}F_{3)}^\gamma-F_\gamma^\beta\nabla_\beta\nabla_{(1}F_{3)}^\gamma\Big]\Big\}
\label{59}
\end{align}
\begin{align}
T^2_{\;3}&=g^{22}\Big\{e^{A}R^2sin\theta\Big(E^{(3)}B^{(2)}-E^{(2)}B^{(3)}\Big)+\frac{a^2}{4\pi}\Big[2F_{(2}^\alpha\nabla_{3)}K_\alpha-2F_{(2}^\alpha\nabla_{\alpha}K_{3)}\Big]\nonumber\\
&+\frac{b^2}{2\pi}\big[F^\gamma_{(2}\nabla_\alpha\nabla_{\beta}F_{3)\gamma}F^{\beta\alpha}-\nabla_{\beta}F_\gamma^\beta\nabla_{(2}F_{3)}^\gamma+\nabla_\beta{F}_\gamma^\beta\nabla_{(2}F_{3)}^\gamma-F_\gamma^\beta\nabla_\beta\nabla_{(2}F_{3)}^\gamma\Big]\Big\}
\label{60}
\end{align}

However, by Eq. $(\ref{5})$, the only non-null off diagonal component of Einstein tensor is $G^1_0$. Therefore, in order to have
\begin{equation}
T_{02}=T_{03}=T_{12}=T_{13}=T_{23}=0
\label{61}
\end{equation} 
we need to impose the conditions
\begin{equation}
F_{02}=F_{03}=F_{12}=F_{13}=F_{23}=0\,\,.
\label{62}
\end{equation}

After imposing these conditions into Eqs. $(\ref{51})-(\ref{60})$ we have that
\begin{align}
T^0_{\;0}=&\frac{1}{2}e^A(E^{(1)})^2+\frac{a^2}{4\pi}\Big[F^{01}H_{01}-\frac{1}{2}K^0K_0+\frac{1}{2}K^1K_1\Big]\nonumber\\
&+\frac{b^2}{2\pi}\Big[-\frac{1}{4}\nabla^{\beta}F^{\alpha\gamma}\nabla_{\beta}F_{\alpha\gamma}+F^{01}\nabla^0\nabla_0F_{01}+F^{01}\nabla^1\nabla_1F_{01}-\nabla^0F_{10}\nabla_0F^{01})\Big]
\label{63}
\end{align}
\begin{align}
T^1_{\;1}=&\frac{1}{2}e^A(E^{(1)})^2+\frac{a^2}{4\pi}\Big[F^{01}H_{01}+\frac{1}{2}K^0K_0-\frac{1}{2}K^1K_1\Big]\nonumber\\
&+\frac{b^2}{2\pi}\Big[-\frac{1}{4}\nabla^{\beta}F^{\alpha\gamma}\nabla_{\beta}F_{\alpha\gamma}+F^{01}\nabla^0\nabla_0F_{01}+F^{01}\nabla^1\nabla_1F_{01}-\nabla^0F_{10}\nabla_0F^{01})\Big]
\label{64}
\end{align}
\begin{align}
T^2_{\;2}=T^3_{\;3}=&\Big[\frac{1}{2}F_{01}F^{01}\Big]+\frac{a^2}{4\pi}\Big[F_0^{\;1}\nabla_1K^0+F_1^{\;0}\nabla_0K^1+K^0K_0+K^1K_1\Big]\nonumber\\
&+\frac{b^2}{2\pi}\Big[-\frac{1}{4}\nabla^{\beta}F^{\alpha\gamma}\nabla_{\beta}F_{\alpha\gamma}\Big]
\label{65}
\end{align}

\begin{align}
T_{10}&=-\Big(\frac{a^2+2b^2}{4\pi}\Big)K_1K_0
\label{67}
\end{align}

So, together with Eq. $(\ref{5})$, we have
\begin{align}
G^0_{\;0}=&\Big(-2\frac{R''}{R}+\frac{R'A'}{R}-\frac{R'^2}{R}\Big)e^{-A}+\Big(\frac{\dot{R}\dot{A}}{R}+\frac{\dot{R}^2}{R^2}+\frac{1}{R^2}\Big)\nonumber\\
&=8{\pi}G\Big\{{\kappa\rho}+\frac{1}{2}e^A(E^{(1)})^2+\frac{a^2}{4\pi}\Big[F^{01}H_{01}-\frac{1}{2}K^0K_0+\frac{1}{2}K^1K_1\Big]\nonumber\\
&+\frac{b^2}{2\pi}\Big[-\frac{1}{4}\nabla^{\beta}F^{\alpha\gamma}\nabla_{\beta}F_{\alpha\gamma}+F^{01}\nabla^0\nabla_0F_{01}+F^{01}\nabla^1\nabla_1F_{01}-\nabla^0F_{10}\nabla_0F^{01})\Big]\Big\}\nonumber\\
\label{68}
\end{align}
\begin{align}
G^1_{\;1}=&-\frac{\dot{R}^2}{R^2}+2\frac{\ddot{R}}{R}-e^{-A}\frac{R'^2}{R^2}+\frac{1}{R^2},\nonumber\\
&=8{\pi}G\Big\{+\frac{1}{2}e^A(E^{(1)})^2+\frac{a^2}{4\pi}\Big[F^{01}H_{01}+\frac{1}{2}K^0K_0-\frac{1}{2}K^1K_1\Big]\nonumber\\
&+\frac{b^2}{2\pi}\Big[-\frac{1}{4}\nabla^{\beta}F^{\alpha\gamma}\nabla_{\beta}F_{\alpha\gamma}+F^{01}\nabla^0\nabla_0F_{01}+F^{01}\nabla^1\nabla_1F_{01}-\nabla^0F_{10}\nabla_0F^{01})\Big]\Big\}
\label{69}
\end{align}
\begin{align}
G^2_{\;2}=&G^3_{\;3}=\Big(-\frac{R''}{R}+\frac{A'R'}{2R}\Big)e^{-A}+\frac{\ddot{R}}{R}+\frac{\ddot{A}}{2}+\frac{\dot{A}^2}{4}-\frac{\dot{R}\dot{A}R}{2}\nonumber\\
&=8{\pi}G\Big[+\frac{1}{2}F_{01}F^{01}\Big]+\frac{a^2}{4\pi}\Big[F_0^{\;1}\nabla_1K^0+F_1^{\;0}\nabla_0K^1+K^0K_0+K^1K_1\Big]\nonumber\\
&+\frac{b^2}{2\pi}\Big[-\frac{1}{4}\nabla^{\beta}F^{\alpha\gamma}\nabla_{\beta}F_{\alpha\gamma}\Big]
\label{70}
\end{align}
\begin{align}
G^1_{\;0}=&e^{-A}\Big(2\frac{\dot{R}'}{R}-\frac{\dot{A}R'}{R}\Big)=-2G\Big(\frac{a^2+2b^2}{4\pi}\Big)K^1K_0.
\label{71}
\end{align}
Hence, in order to obtain a more concise notation, let us define
\begin{equation}
\Big(\frac{a^2+2b^2}{4\pi}\Big){\equiv}M^2\,\,.
\end{equation}

Hence, that Eq. $(\ref{71})$ can be written as
\begin{equation}
e^{-A}\Big(2\frac{\dot{R}'}{R}-\frac{\dot{A}R'}{R}\Big)=-2GM^2K^1K_0.
\label{71.1}
\end{equation}
where, together with the definition of covariant derivative in Eq. $(\ref{6.1})$ we have that
\begin{align}
K_0=&\partial_1\Big(\frac{1}{2}A(t,r)-E(t,r)\Big)\,\,,\nonumber\\
K_1=&\partial_0\Big(\frac{1}{2}A(t,r)-E(t,r)\Big)\,\,.
\label{72}
\end{align}

Integrating Eq. $(\ref{71.1})$, we obtain
\begin{equation}
R'=e^{A/2}\chi(t,r)\,\,,
\label{72.1}
\end{equation}
where
\begin{equation}
\chi(t,r)=-{\int}2GM^2K^1K_0dt+\alpha(r)\,\,,
\label{72.2}
\end{equation}
where $\alpha$ is the constant of integration and, as we can see, if the $M$ is zero, we obtain the Maxwell case discussed in \cite{Fanizza}. Therefore, defining $\chi\,{\equiv}\,1-k(t,r)$, we obtain
\begin{equation}
e^A=\frac{R'^2}{(1-k(t,r))}
\label{72.3}
\end{equation}
and the line element in Eq. $(\ref{1})$ can be write as
\begin{equation}
dS^2=dt^2-\frac{R'^2}{(1-k(t,r))}dr^2-R^2\Big(d\theta^2+sin^2{\theta}d\phi^2\Big)\,\,.
\label{72.4}
\end{equation}
The new result here is that the $k$-term  depends on both $r$ and $t$
coordinate.

For the Podolsky-Maxwell Eqs. $(\ref{46})$, with the conditions in Eqs. $(\ref{51})-(\ref{60})$, the non-zero components are
\begin{align}
F_{01}-(a^2+2b^2)H_{01}+2b^2S_{01}&=C(t)\frac{e^{A/2}}{R^2}\nonumber\\
F_{10}-(a^2+2b^2)H_{10}+2b^2S_{10}&=D(r)\frac{e^{A/2}}{R^2}
\label{73}
\end{align}
where clearly, 
\begin{equation}
C(t)=-D(r)
\label{74}
\end{equation}
and 
\begin{align}
F_{01}-(a^2+2b^2)H_{01}+2b^2S_{01}&=C\frac{e^{A/2}}{R^2}
\label{75}
\end{align}

The line element in Eq. $(\ref{72.4})$ together with Eq. $(\ref{47})$ lead us to 
\begin{align}
S_{10}&=-F_{01}\bigglb[2\frac{\ddot{R}}{R}-\frac{\dot{R}\dot{A}}{R}-\Big(\frac{R'A'}{R}-2\frac{R''}{R}\Big)e^{-A}\biggrb]\nonumber\\
\label{76}
\end{align}

Therefore, using Eq. $(\ref{72.4})$, we rewrite the Einstein equations for the $(00)$ and $(11)$ components as 
\begin{align}
G^0_{\;0}=&\frac{\dot{R}^2}{R^2}+2\frac{\dot{R}\dot{R'}}{RR'}-\frac{\dot{R}\dot{k}}{R(1-k)}+\frac{k}{R^2}+\frac{kk'R'}{R^3(1-k)}\nonumber\\
&=8{\pi}G\Big\{\kappa\rho+\frac{R'^2}{2(1+2k(t,r))}(E^{(1)})^2+\frac{a^2}{4\pi}\Big[F^{01}H_{01}-\frac{1}{2}K^0K_0+\frac{1}{2}K^1K_1\Big]\nonumber\\
&+\frac{b^2}{2\pi}\Big[-\frac{1}{4}\nabla^{\beta}F^{\alpha\gamma}\nabla_{\beta}F_{\alpha\gamma}+F^{01}\nabla^0\nabla_0F_{01}+F^{01}\nabla^1\nabla_1F_{01}-\nabla^0F_{10}\nabla_0F^{01})\Big]\Big\}\nonumber\\
\label{77}
\end{align}
\begin{align}
G^1_{\;1}=&-\frac{\dot{R}^2}{R^2}+2\frac{\ddot{R}}{R}-(1+2k(t,r)){R^2}+\frac{1}{R^2}\nonumber\\
&=8{\pi}G\Big\{\frac{R'^2}{2(1-k(t,r))}(E^{(1)})^2+\frac{a^2}{4\pi}\Big[F^{01}H_{01}+\frac{1}{2}K^0K_0-\frac{1}{2}K^1K_1\Big]\nonumber\\
&\frac{b^2}{2\pi}\Big[-\frac{1}{4}\nabla^{\beta}F^{\alpha\gamma}\nabla_{\beta}F_{\alpha\gamma}+F^{01}\nabla^0\nabla_0F_{01}+F^{01}\nabla^1\nabla_1F_{01}-\nabla^0F_{10}\nabla_0F^{01})\Big]\Big\}
\label{78}
\end{align}

Defining, 
\begin{equation}
\epsilon(t,r)\,{\equiv}\,\frac{1}{2}K^1K_1-\frac{1}{2}K^0K_0+F^{01}\nabla^0\nabla_0F_{01}-F^{01}\nabla^1\nabla_1F_{01}
\end{equation}
and
\begin{equation}
\epsilon'(t,r)\,{\equiv}\,\frac{1}{2}K^1K_1-\frac{1}{2}K^0K_0-F^{01}\nabla^0\nabla_0F_{01}+F^{01}\nabla^1\nabla_1F_{01}\,\,,
\end{equation}

\ni we can rewrite these Einstein equations as
\begin{align}
&\frac{\dot{R}^2}{R^2}+2\frac{\dot{R}\dot{R'}}{RR'}-\frac{\dot{R}\dot{k}}{R(1-k)}+\frac{k}{R^2}+\frac{kk'R'}{R^3(1-k)}\nonumber\\
&=8{\pi}G\bigglb({\kappa\rho}+\frac{R'^2}{2(1-k(t,r))}(E^{(1)})^2+M^2\epsilon'(t,r)\biggrb)
\label{78.1}
\end{align}
and
\begin{align}
&\frac{\dot{R}^2}{R^2}+2\frac{\ddot{R}}{R}+k(t,r){R^2}
=8{\pi}G\bigglb\{\frac{R'^2}{2(1-k(t,r))}(E^{(1)})^2+M^2\epsilon(t,r)\biggrb\}
\label{79}
\end{align}

\ni which lead us to 
\begin{equation}
\dot{R}^3+2R\dot{R}\ddot{R}=k(t,r)\dot{R}+\sigma(t,r)\frac{\dot{R}}{R^2}
\label{80}
\end{equation}
where 
\begin{equation}
\sigma(t,r)\,{\equiv}\,4{\pi}Ge^AR^4\Big(E^2+2M^2\epsilon(t,r){e^A}\Big)
\label{81}
\end{equation}

So, we write $\dot{R}$ as
\begin{equation}
\dot{R}=\frac{1}{\sqrt{R}}{\int}K(t,r)dt+\frac{1}{\sqrt{R}}{\int}\sigma(t,r)dt+\frac{\alpha(r)}{\sqrt{R}}
\label{82}
\end{equation}
substituting this result into Eq. $(\ref{78.1})$ we obtain
\begin{align}
&8{\pi}G\Big\{{\kappa\rho+\frac{R'^2}{2(1-k(t,r))}(E^{(1)})^2-M^2\epsilon'(t,r)
}\Big\}\nonumber\\
&=\,\frac{2}{RR'}\bigglb(\Sigma(t,r)+\Omega(t,r)+\frac{\alpha}{\sqrt{R}}\bigglb)\biggrb[\Sigma'(t,r)+\Omega'(t,r)+\partial_r\bigglb(\frac{\alpha}{\sqrt{R}}\biggrb)\biggrb]\nonumber\\
&+\frac{1}{R^2}\bigglb(\Sigma(t,r)+\Omega(t,r)+\frac{\alpha}{\sqrt{R}}\biggrb)^2\nonumber\\
&+\frac{1}{R}\bigglb(\Sigma(t,r)+\Omega(t,r)+\frac{\alpha}{\sqrt{R}}\biggrb)\frac{\dot{k}}{R(1-k)}+\frac{k}{R^2}+\frac{kk'R'}{R^3(1-k)}
\label{83}
\end{align}

Eqs. $(\ref{72.4})$, $(\ref{82})$ and $(\ref{83})$ define the LTB model with the Podolsky electrodynamics contributions. The singularities arising from $R=0$, $R'=0$ and $k=1$. 
The $R=0$ singularity can be understood as the Big Bang singularity, and the $R'=0$ singularity as the shell cross singularity. The last singularity, $k=1$, comes from the Podolsky contribution. 

\section{Conclusion}

The analysis of the electromagnetic field in an LTB background was explored
by other authors in Ref.\ \cite{Fanizza}, however they studied only the Maxwell
case, which means that the effects of a mass term or the complication of
high-derivative electromagnetic term were not investigated until now. The
objective of this paper is to fill this gap and to analyze the possible
theoretical effects of these rich electrodynamic scenarios in a more
realistic inhomogeneous cosmological background.

In this work, we have analyzed the Proca and the higher-derivative Podolsky
models embedded in an inhomogeneous LTB background. In the case of Proca
model we found a new singularity at $k(r)=0$. Moreover, the magnetic field
must be zero to satisfy the Einstein's equations. Considering the Proca
model, we have determined the scale factor and provided an analysis of the
luminosity distance. 

In the Podolsky case, we analyzed the function $k$, which appears in the
line element that defines the LTB model where the Podolsky electrodynamics
is dependent of both $t$ and $r$ coordinate. This is different from Maxwell
and Proca cases.

As a perspective, a more precise analysis of the Podolsky contributions for
the cosmological LTB model can be carried out, as the analysis of black
holes to generate the singularity at $R'=0$, for example.

\section*{Acknowledgments}

\ni E.M.C.A. thanks CNPq (Conselho Nacional de Desenvolvimento 
Cient\'ifico e Tecnol\'ogico), Brazilian scientific support 
federal agency, for partial financial support, Grants numbers 
406894/2018-3 and 302155/2015-5.


\begin{thebibliography}{99}

\bibitem{Pascual}
J. F. Pascual-Sanchez, ``Cosmic acceleration: inhomogeneity versus vacuum energy," Mod. Phys. Lett. A 14 (1999) 1539.

\bibitem{Rasanen}
S. Rasanen, `` Accelerated expansion from structure formation," JCAP 11 (2006) 003.

\bibitem{Chuang} 
C. H. Chuang, J. A. Gu and W. Y. P. Hwang, ``Inhomogeneity-induced cosmic acceleration in a dust universe," Class. Quant. Grav. 25 (2008) 175001.

\bibitem{Paranjape}
A. Paranjape and T.P. Singh, ``The possibility of cosmic acceleration via spatial averaging in LTB models," Class. Quant. Grav. 23 (2006) 6955.

\bibitem{Kai} 
T. Kai, H. Kozaki, K. I. Nakao, Y. Nambu and C. M. Yoo, ``Can inhomogeneities accelerate the cosmic volume expansion?," Prog. Theor. Phys. 117 (2007) 229.

\bibitem{Rasanen2}
S. Rasanen, ``Cosmological acceleration from structure formation," Int. J. Mod. Phys. D 15 (2006) 2141.

\bibitem{Enqvist} 
K. Enqvist, ``LTB model and acceleration expansion," Gen. Relat. Grav. 40 (2008) 451.

\bibitem{Cosmai} 
L. Cosmai, G. Fanizza, M. Gasperini and L. Tedesco, ``Discrimination different models of luminosity-redshift distribution," Class. Quant. Grav. 30 (2013) 095011.

\bibitem{Bondi}
W. B. Bonnor, ``The formation of nebulae," Zeitschrift f\"ur Astrophysik, 39 (1956) 143.

\bibitem{Krasinski}
A. Krasinski, {\it Inhomogeneous Cosmological Models}, Cambridge University Press, CUP, 1997.

\bibitem{Lemaitre}
G. Lema\^{\i}tre,  ``L'Universe en expasion," Ann. Soc. Scient. Bruxelles A 53 (1933) 51.

\bibitem{Ribeiro}
M. B. Ribeiro,  ``On Modelling a Relativistic Hierarchical (Fractal) Cosmology Tolman's Spacetime. I. Theory," Astrophys. J., 388 (1992) 1, arXiv: 0807.0866.

\bibitem{Ribeiro1}
M. B. Ribeiro, ``On Modelling a Relativistic Hierarchical (Fractal) Cosmology Tolman's Spacetime. II. Analysis of Einstein-de Sitter model," Astrophys. J., 395 (1992) 29, arXiv: 0807.0869.

\bibitem{Ribeiro2}
M. B. Ribeiro, ``On Modelling a Relativistic Hierarchical (Fractal) Cosmology Tolman's Spacetime. III. Numerical results," Astrophys. J., 415 (1993) 469, arXiv: 0807.1021.

\bibitem{Ribeiro3}
M. B. Ribeiro, ``Relativistic Fractal Cosmologies,"  NATO Sci. Ser. B 332 (1994) 269, arXiv: 0910.4877.

\bibitem{Ribeiro4}
M. B. Ribeiro, ``The apparent Fractal Conjecture: Scaling Features in Standard Cosmologies," Gen. Relativ. Gravit., 33 (2001) 1699, arXiv: astro-ph/0104181.

\bibitem{Nogueira}
F. A. M. G. Nogueira, ``Single past null geodesic in the LTB cosmology," M.Sc. Dissertation, arXiv: 1312.5005.

\bibitem{ciarcelluti}   
P. Ciarcelluti, ``Electrodynamic effect of anisotropic expansions in the Universe," Mod. Phys. Lett. A 27 (2012) 1250221.

\bibitem{Fanizza}
G. Fanizza and L. Tedesco, ``Electrodynamics in an LTB scenario," Eur. Phys. J. C 74 (2014) 2786.

\bibitem{ybr}   
Z. Yousaf, M. Z. Bhatti and A. Rafaqat, ``Electromagnetic effects on the evolution of LTB geometry in modified gravity," 
Astrophys. Space Sci. 362 (2017) 68.

\bibitem{Bekenstein}
J. D. Bekenstein, ``Nonexistence of baryon number for static black holes," Phys. Rev. D 5 (1972) 1239.
 
\bibitem{Cuzinatto}
R. R. Cuzinatto, C. A. M de Melo, L. G. Medeiros, B. M. Pimentel and P. J. Pompeia, ``Bopp-Podolsky black holes and no-hair theorem," Eur. Phys. J. C 78 (2018) 43.

\end{thebibliography}
\end{document}